\documentclass[12pt]{iopart}
\usepackage{graphicx}
\begin{document}

\title[Abundances and Ultra-Low Momentum Neutrons]{Nuclear Abundances in Metallic Hydride Electrodes \\ 
of Electrolytic Chemical Cells}

\author{A. Widom\dag\ and L. Larsen\ddag\ }

\address{\dag\ Physics Department, Northeastern University\\ 
110 Forsyth Street, Boston MA 02115}

\address{\ddag\ Lattice Energy LLC, 175 North Harbor Drive, Chicago IL 60601}

\begin{abstract}
Low energy nuclear transmutations have been reported in experimental chemical electrolytic 
cells employing metallic hydride electrodes. Assuming that the nuclear transmutations  
are induced by ultra-low momentum neutron absorption, the expected chemical cell nuclear abundances 
are discussed on the basis of a neutron optical potential model. The theoretical results 
are in satisfactory agreement with available experimental chemical cell data. Some implications 
of these laboratory nuclear transmutations for r- and s-process models of the neutron induced solar system 
and galactic nuclear abundance are briefly explored.
\end{abstract}



\section{Introduction \label{Intro}}

It has been theoretically asserted\cite{Widom:2005}, for surfaces of fully 
loaded metallic hydrides saturated with protons, that the collective electron and 
proton surface plasma modes give rise to the production of ultra-low momentum neutrons. 
The reaction proceeds in accordance with the weak interaction model,
\begin{eqnarray} 
({\rm radiation\ energy})+e^- \to \tilde{e}^{\ -},
\nonumber \\ 
\tilde{e}^{\ -}+p^+\to  n+\nu_e ,
\label{Intro1} 
\end{eqnarray} 
wherein \begin{math} e^- \end{math} is an electron, 
\begin{math} \tilde{e}^{\ -} \end{math} is a  
``dressed'' or heavy electron localized near the metal hydride surface, 
\begin{math}  p^+  \end{math} is a proton, \begin{math}  n  \end{math} 
is a neutron and \begin{math}  \nu_e  \end{math} is an electron neutrino. 
The resulting hydride surface product ultra-low momentum 
neutrons have an extraordinarily high absorption cross section.
These neutrons in turn yield successive 
nuclear transmutations into higher and higher values of the atomic mass number 
\begin{math} A \end{math}. At appropriate values of \begin{math} A \end{math},  
a resulting unstable nucleus may beta decay so that the values of 
the nuclear charge number \begin{math} Z \end{math} then also rises. 
From such energy nuclear reaction kinetics, most of the periodic 
table of chemical elements may be produced, at least to some extent. 
Such nuclear transmutations do not require the high Coulomb energy barrier 
penetration implicit in low energy fusion models.  

It has been previously reported\cite{Miley:2005}, for experimental electrolytic chemical 
cells employing metallic hydride electrodes, that a variety of 
nuclear transmutations have indeed been observed and that the resulting  
abundances of various nuclei have been reliably measured. Our purpose is to 
describe the experimental abundances of nuclear transmutations in strongly 
driven electro-chemical cells in terms of theoretical ultra-low momentum 
neutron absorption cross sections. 

In Sec.\ref{potential}, a simple neutron 
optical potential model for a spherical well is reviewed 
and the complex scattering lengths for describing the ultra-low momentum 
neutron absorption cross sections are derived\cite{Landau:1977}.  
In Sec.\ref{transmutation}, we discuss the experimental distribution in atomic mass number 
\begin{math} A \end{math} of the low energy nuclear reaction products measured 
in laboratory chemical cells\cite{Miley:1997a}. Very remarkably, the product yield in a chemical cell 
is in some ways qualitatively similar to nuclear abundances found in our local solar 
system and galaxy\cite{Sneden:2003}. The local maxima and minima in these abundances  are strongly 
correlated to the local maxima and minima in the ultra-low momentum neutron absorption cross sections. 
This strong similarity serves as an indicator of the potential importance of ultra-low energy 
neutrons in nuclear reaction kinetics as briefly discussed in the concluding Sec.\ref{conc}.

\section{Optical Potential Theory \label{potential}}

\begin{figure}[tp]
\scalebox {0.7}{\includegraphics{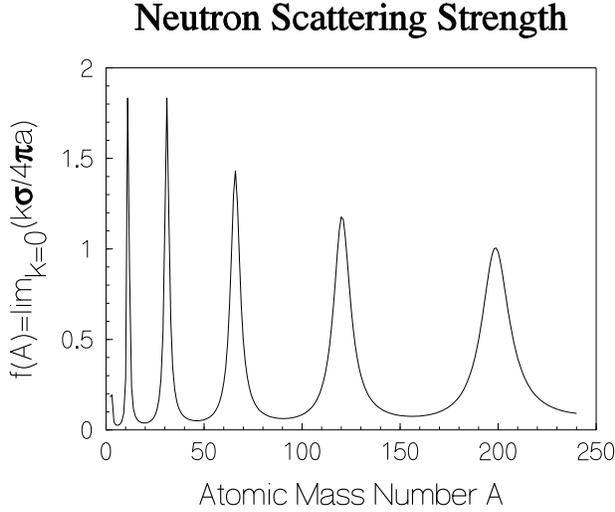}}
\caption{In the ultra-low momentum limit $k\to 0$, the total cross 
section is given by $\sigma (k,A)=(4\pi {\rm a}/k)f(A)$. The scattering 
strength $f(A)$ is plotted as function of atomic mass number $A$.}
\label{fig1}
\end{figure}

Let us consider the following {\em optical potential model} for the effective (added to a nucleus) 
neutron amplitude \begin{math} \psi ({\bf r})  \end{math}. The complex potential probed by the neutron 
has the form of a spherical well. The well radius for a given atomic mass number 
\begin{math} A \end{math} is modeled by 
\begin{equation}
R=aA^{1/3}\ \ \ {\rm wherein}\ \ \ a=1.2\times 10^{-13}\ {\rm cm}.
\label{potential1}
\end{equation}
In detail, we assume a complex spherical step potential well 
\begin{eqnarray}
U(r<R) &=& -\left(V+\frac{i\hbar \Gamma}{2}\right),
\nonumber \\ 
U(r>R) &=& 0.
\label{potential2}
\end{eqnarray}
From the Scr\"odinger equation for the neutron amplitude 
\begin{equation}
\left(-\frac{\hbar ^2}{2M}\Delta +U(r)\right)
\psi ({\bf r})=E\psi ({\bf r})=\frac{\hbar ^2k^2}{2M}\psi ({\bf r}),
\label{potential3}
\end{equation}
one finds the scattering amplitude \begin{math} {\cal F}(k,\theta ) \end{math},
\begin{equation}
\psi ({\bf r})\to e^{ikz}+{\cal F}(k,\theta )\frac{e^{ikr}}{r}+\ldots 
\ \ {\rm as}\ \ r\to \infty,
\label{potential4}
\end{equation}
and thereby the total neutron cross section 
\begin{equation}
\sigma (k)=\frac{4\pi }{k}{\Im m}{\cal F}(k,0).
\label{potential5}
\end{equation}
In the ultra-low momentum limit, one may compute the neutron 
scattering strength 
\begin{equation}
f(A)=\lim_{k\to 0}\left(\frac{k\sigma (k)}{4\pi a}\right)
=\lim_{k\to 0}\left(\frac{{\Im m}{\cal F}(k,0)}{a}\right).
\label{potential6}
\end{equation}
The exact analytic form of the limiting Eq.(\ref{potential6}) is given by  
\begin{eqnarray}
f(A) = {\Im m}\left\{\frac{\tan\big(zA^{1/3}\big)}{z}\right\},
\nonumber \\ 
\frac{\hbar z}{\rm a} = \sqrt{2M\left(V+i\frac{\hbar \Gamma}{2}\right)}\ .
\label{potential7}
\end{eqnarray}
Employing this optical potential model with \begin{math}z=3.5+0.05i \end{math}, 
the total neutron cross section in the long wavelength limit may be analytically computed,   
\begin{equation}
\sigma (k,A)=\left(\frac{4\pi {\rm a}}{k}\right)f(A)
\ \ \ {\rm as}\ \ \  k\to 0,
\label{potential8}
\end{equation}
as plotted in Fig.\ref{fig1}. The peaks in the neutron cross section 
correspond to comfortably fitting the neutron wave within the spherical 
model optical potential wells of the nuclei. The well radius 
\begin{math} R  \end{math} varies with \begin{math} A  \end{math} in 
accordance with Eq.(\ref{potential1}).

While the model solved above refers the absorption of one neutron by one 
spherical nucleus, when the model is applied an ultra-low momentum neutron 
in a condensed matter electrode of a chemical cell, the very long neutron 
wave length spans many nuclei. This gives rise to collective quantum 
coherent effects in the neutron-nuclei interaction. 

\section{Nuclear Reactions in a Chemical Cell \label{transmutation}}

\begin{figure}[tp]
\scalebox {0.8}{\includegraphics{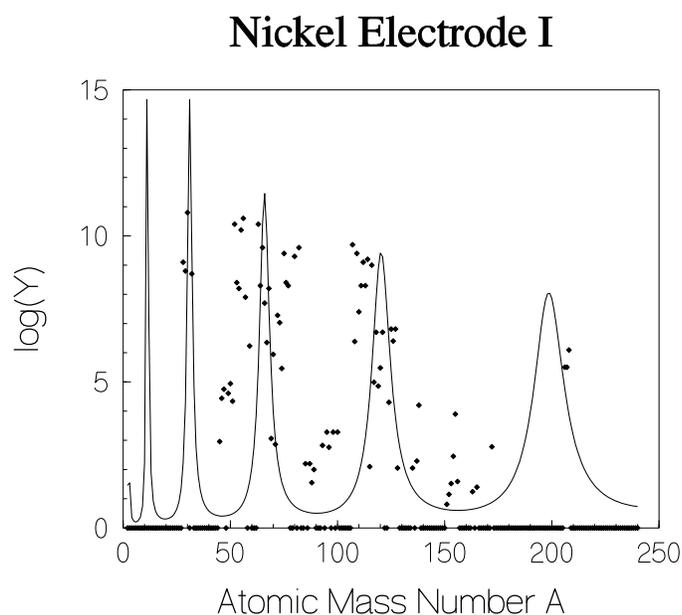}}
\caption{Shown is the experimental yield $Y$ in parts per million 
Nickel electrode atoms of nuclear transmutation products during a moderately 
productive run.  The experimental points were produced employing chemical electrolytic 
cells with a Nickel hydride cathode. Points exactly on the $A$-axis are those below 
detectable experimental resolution. Also plotted is a smooth theoretical curve 
of $8\times f(A)$ which is proportional to the neutron absorption cross section.}
\label{fig2}
\end{figure}

\begin{figure}[bp]
\scalebox {0.8}{\includegraphics{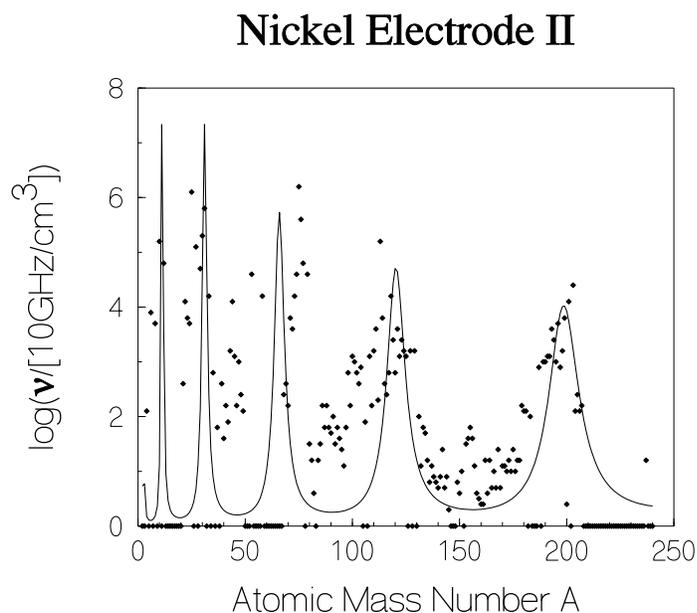}}
\caption{Shown is the experimental yield $\nu$ measured in transmutation 
events per second per cubic centimeter of Nickel electrode during a  
very productive run. The experimental points were produced employing chemical 
electrolytic cells with a Nickel hydride cathode. Points exactly on the 
$A$-axis are those below experimental resolution. Also plotted is a smooth 
theoretical curve of $4\times f(A)$ which is proportional to the neutron 
absorption cross section.}
\label{fig3}
\end{figure}

The following experimental nuclear transmutations have been previously reported 
using light water solutions of one Molar \begin{math} Li_2 SO_4  \end{math} in 
electrolytic cells. Small plastic micro-spheres were coated with thin films of Nickel 
which were then employed to construct electrodes. The granular thin film structure of the 
electrodes allowed for large effective surface areas for the nickel hydride. 
If there exists sufficiently large flux of protons moving through the nickel and if the 
surface of the nickel hydride film is saturated with hydrogen atoms and thereby protons, 
then nuclear transmutations could be measured to take place with a yield that is plotted 
as functions of \begin{math} A \end{math}. The magnitude of the transmuted nuclear yields 
varies from one experimental run to to another. The variations will depend on the isotopic 
composition of the metallic hydride cathode and anode, various ions found in solution and 
the nano-scale uniformity of the electrode fabrication processes. However, the distribution of the 
yields versus atomic number appear to be understandable in terms of the total cross sections 
for neutron absorption in the ultra-low momentum limit, i.e. in terms of the above 
calculated neutron scattering strength. 

In Fig.\ref{fig2}, we plot the reported transmutation yield versus atomic mass number 
for a moderately productive run\cite{Miley:1996} employing a Nickel cathode and compare 
the observed shape of the yield function and the shape of the ultra-low momentum neutron 
scattering strength. In Fig.\ref{fig3} we plot the reported transmutation yield versus 
atomic mass number for a very productive run\cite{Miley:1997b} employing a Nickel cathode 
and again compare the observed shape of the yield function and the shape of the ultra-low 
momentum neutron scattering strength. Similar experimental 
results\cite{Miley:2005}\cite{Miley:1997a}\cite{Miley:1996}\cite{Miley:1997b} have been found 
employing titanium hydride, palladium hydride, and layered Pd-Ni metallic hydride electrodes. 
In all such experimental runs, the agreement between the multi-peak transmutation yields 
and the neutron scattering strength is quite satisfactory.

\section{Conclusions \label{conc}}

Let us conclude with some comments on the nature of the peaks 
in the theoretical neutron scattering strengths shown in Fig.\ref{fig1}.
In varying the atomic mass number \begin{math} A \end{math}, we are in 
reality varying the radius \begin{math} R=aA^{1/3} \end{math} of the optical 
potential well. When the neutron wavelength within the well reaches  
resonance with the radius of the well\cite{Sill:2005} a peak appears in the 
scattering strength. If we associate resonant couplings with the ability of 
the neutron to be virtually trapped in a region neighboring the nucleus, then 
for intervals of atomic mass numbers about and under the resonant peaks we would 
expect to obtain neutron halo nuclei\cite{Riisager:1994}\cite{Jensen:2004}. 
The spherical optical potential 
well\cite{Liu:2003} predicts the halo nuclei stability valleys\cite{Vogt:2002} and 
thus the peaks in observed nuclear transmutation abundances. Finally, the neutrons 
yielding the abundances in our local solar system and galaxy have often been previously 
assumed to arise entirely from stellar nucleosynthetic processes and supernova explosions. 
However, such assumptions may presently be regarded as suspect\cite{Woosley:2005}\cite{Kratz:1993}. 
It appears entirely possible that ultra-low momentum neutron absorption may have an important 
role to play in the nuclear abundances not only in chemical cells but also in our local 
solar system and galaxy.

\end{document}